\documentclass[1p, final]{elsarticle}

\usepackage{subfig}
\usepackage{amsmath,bm}
\usepackage{color}
\usepackage{enumerate}
\usepackage{multirow}
\usepackage{amssymb}
\usepackage{epstopdf}
\usepackage{epsfig}
\usepackage[table,xcdraw]{xcolor}
\usepackage{algpseudocode,algorithm,algorithmicx}

\usepackage{tikz}
\usepackage{graphicx}
\usetikzlibrary{positioning}

\usepackage{bigstrut}
\usepackage{lineno,hyperref}
\modulolinenumbers[5]

\journal{ }

\begin{document}

\begin{frontmatter}

\title{Graph Convolutional Networks for Simulating Multi-phase Flow and Transport in Porous Media}

\author{Jiamin Jiang}
\cortext[mycorrespondingauthor]{Corresponding author}
\ead{jiangjiaxz@outlook.com}

\address{Chevron Technical Center}

\author{Bo Guo}

\address{Hydrology and Atmospheric Sciences, The University of Arizona}

\date{July 10, 2023}

\begin{abstract}

Numerical simulation of multi-phase fluid dynamics in porous media is critical for many energy and environmental applications in Earth's subsurface. Data-driven surrogate modeling provides computationally inexpensive alternatives to high-fidelity numerical simulators. While the commonly used convolutional neural networks (CNNs) are powerful in approximating partial differential equation solutions, it remains challenging for CNNs to handle irregular and unstructured simulation meshes. However, simulation models for Earth's subsurface often involve unstructured meshes with complex mesh geometries, which limits the application of CNNs. To address this challenge, we construct surrogate models based on Graph Convolutional Networks (GCNs) to approximate the spatial-temporal solutions of multi-phase flow and transport processes in porous media. We propose a new GCN architecture suited to the hyperbolic character of the coupled PDE system, to better capture transport dynamics. Results of 2D heterogeneous test cases show that our surrogates predict the evolutions of pressure and saturation states with high accuracy, and the predicted rollouts remain stable for multiple timesteps. Moreover, the GCN-based models generalize well to irregular domain geometries and unstructured meshes that are unseen in the training dataset.

\end{abstract}

\end{frontmatter}

\section{Introduction}

Dynamics of multiple fluid phases in porous media are critical for many applications in Earth's subsurface, including oil and gas recovery, groundwater remediation, geological $\mathrm{CO_2}$ sequestration, and subsurface hydrogen storage. Numerical simulations play an increasingly important role in understanding, quantifying, and controlling these multi-phase flow processes. Predicting the evolution of subsurface fluid dynamics requires solving partial differential equations (PDEs) governing the multi-phase flow and transport processes. These PDEs are often highly nonlinear and exhibit an intricate mixture of elliptic and hyperbolic characteristics, posing challenges to numerical methods. Moreover, significant uncertainties are present in the model parameters due to data scarcity in the subsurface. As a result, many model simulation runs (e.g., thousands) are required to quantify the uncertainties propagated from the parameters to the predictions. Therefore, computationally efficient simulation techniques are critical for applications in Earth's subsurface.


The deep learning revolution (LeCun et al. 2015; Krizhevsky et al. 2017) has dramatically changed scientific fields such as computer vision and natural language processing. More recently, deep learning algorithms have been extended towards constructing data-driven surrogate models to approximate the solutions of PDEs, particularly in the context of fluid dynamics (Guo et al. 2016; Kutz 2017; Long et al. 2018; Bar-Sinai et al. 2019; Santos et al. 2020; Li et al. 2020; Wang et al. 2021; Lu et al. 2021; Vinuesa and Brunton 2022). Compared to high-fidelity numerical simulators, a learned simulator can provide much faster predictions, especially for high-dimensional nonlinear systems.

A number of studies have applied image-based approaches and snapshots of simulation data over a spatially discretized input domain for surrogate modeling of subsurface flow and transport problems. Most of these works leverage convolutional neural networks (CNNs) to learn the nonlinear mappings from the input properties (e.g., permeability) to the output states (pressure and saturation of fluids) on regular Cartesian meshes (Mo et al. 2019; Tang et al. 2020; Wang and Lin 2020; Wen et al. 2021; Zhang et al. 2021; Jiang et al. 2021; Yan et al. 2022; Maldonado-Cruz and Pyrcz 2022). While CNNs are powerful in approximating PDE solutions, they are restricted to a specific discretization of the physical domain in which they are trained. Due to the inherent limitations of standard convolution operations, it remains challenging for CNNs to handle irregular and unstructured simulation meshes. However, driven by the need to accurately characterize complex geological features and heterogeneity, subsurface simulation models often involve corner-point and unstructured meshes with skewed and degenerate mesh geometries. These complexities limit the application of CNN-based models for subsurface problems. Note that Maucec and Jalali (2022) recently applied the interaction networks (Battaglia et al. 2016) for surrogate modeling of a two-phase incompressible flow problem, but the resulting surrogate leads to large prediction errors of the pressure field under certain conditions.


Graph Neural Networks (GNNs) have successfully been employed to learn the dynamic evolutions of PDEs, under mesh-based simulation frameworks (Pfaff et al. 2020; Belbute-Peres et al. 2020; Iakovlev et al. 2020; Chen et al. 2021; Brandstetter et al. 2022; Pilva and Zareei 2022). In contrast to CNNs, GNNs naturally enable operating on unstructured meshes with complex domain boundaries. A simulation mesh can be viewed as a graph composed of nodes, and a set of edges representing the connectivity between the nodes. The key idea of GNNs is to aggregate and propagate the local information of system states from their neighborhoods into node representations, through multiple message passing layers (Kipf and Welling 2016; Gilmer et al. 2017).


In the present work, we apply Graph Convolutional Networks (GCNs) to learn surrogate models for predicting the spatial-temporal solutions of multi-phase flow and transport in porous media. We seperately design two GCN architectures that are suited to the elliptic and hyperbolic characteristics of the coupled PDE system, to better capture the dynamics of fluid pressure and saturation. The GCN-based models are trained by supervising on the per-node output states. We evaluate the prediction performance of the trained surrogates using 2D heterogeneous cases. The results show that our surrogates predict the dynamic evolutions with high accuracy, and the predicted rollouts remain stable for multiple timesteps. Moreover, our GCN models generalize well to irregular domain geometries and unstructured meshes that are not present in the training dataset.

\section{Mathematical model and discretization}

\subsection{Immiscible multi-phase flow in porous media}

We consider compressible and immiscible flow and transport in porous media with $n_p$ number of phases. The mass-conservation equation for phase $l$ ($l \in \left \{ 1,...,n_p \right \}$) can be written as
\begin{equation} 
\label{eq:mass_con}
\frac{\partial }{\partial t } \left ( \phi \rho_{l} s_{l} \right ) + \nabla \cdot \left (\rho_{l} \textbf{\textit{v}}_{l} \right ) - \rho_{l} q_{l} = 0,
\end{equation}
where $t$ is time. $\phi$ is rock porosity. $q_{l}$ is the volumetric injection or pumping rate of wells (source or sink term). $\rho_{l}$ is phase density. $s_{l}$ is fluid phase saturation, which is constrained by
\begin{equation} 
\sum_{l} s_{l} = 1,
\label{eq:s_con}
\end{equation}

The Darcy phase velocity, $\textbf{\textit{v}}_l$, is expressed as
\begin{equation} 
\label{eq:phase_vel}
\textbf{\textit{v}}_l = -k \lambda_l \left ( \nabla p_l - \rho_l g \nabla z \right ).
\end{equation}
where $k$ is rock permeability. $p_l$ is phase pressure. $g$ is gravitational acceleration and $z$ is depth (assuming positive downward). $\lambda_l = k_{rl}/\mu_l$ is phase mobility, where $k_{rl}$ and $\mu_l$ are relative permeability and fluid viscosity, respectively.

For a system that only involves two fluid (nonwetting and wetting) phases, Eq.~(\ref{eq:mass_con}) can be simplified to
\begin{equation} 
\frac{\partial}{\partial t } \left ( \phi \rho_{nw} s_{nw} \right ) + \nabla \cdot \left ( \rho_{nw} \textbf{\textit{v}}_{nw} \right ) - \rho_{nw} q_{nw} = 0, 
\end{equation}

\begin{equation} 
\frac{\partial}{\partial t } \left ( \phi \rho_{w} s_{w} \right ) + \nabla \cdot \left ( \rho_{w} \textbf{\textit{v}}_{w} \right ) - \rho_{w} q_{w} = 0, 
\end{equation}
with the saturation constraint as $s_{nw} + s_w - 1 = 0$. The corresponding capillary pressure is defined as the difference in phase pressures
\begin{equation}
p_c = p_{nw} - p_{w}.
\end{equation}

\subsection{Fully-implicit discretization}

To solve the PDE system from Eq.~(\ref{eq:mass_con}), we apply a finite volume method that discretizes the simulation domain into a mesh consisting of $n_b$ cells and a fully-implicit scheme for the time discretization
\begin{equation} 
\label{eq:dis_mass}
\frac{\left | \Omega_i \right |}{\Delta t} \left ( \left ( \phi_i \rho_{l,i} s_{l,i} \right )^{n+1} - \left ( \phi_i \rho_{l,i} s_{l,i} \right )^{n} \right ) - \sum_{j\in adj(i)} \! \left ( \rho_{l,ij} \upsilon_{l,{ij}} \right )^{n+1} - Q_{l,i}^{n+1} = 0,
\end{equation}
where $i \in \left \{ 1,...,n_b \right \}$ is cell index, $\left | \Omega_i \right |$ is cell volume, $(ij)$ corresponds to the interface between cells $i$ and $j$. Superscripts represent timesteps, and $\Delta t$ is timestep size.

The discrete phase flux based on the two-point flux approximation can be written as
\begin{equation} 
\label{eq:dis_p_f}
\upsilon_{l,ij} = T_{ij} \lambda_{l,ij} \Delta \Phi_{l,ij},
\end{equation}
where $\Delta \Phi_{l,ij} = \Delta p_{l,ij} - g_{l,ij}$ is the phase-potential difference with the discrete weights $g_{l,ij} = \rho_{l,ij} \, g \Delta z_{ij}$. The phase mobility $\lambda_{l,ij}$ is evaluated using the Phase-Potential Upwinding (PPU) scheme (Sammon 1988; Brenier and Jaffré 1991). In PPU, the mobility of each phase is treated separately according to the sign of the phase-potential difference. The upwinding criterion is given as
\begin{equation} 
\label{eq:PPU}
\lambda_{l,ij} = \left\{ {\begin{array}{*{20}c}
\lambda_{l}(s_{i}),   & \Delta \Phi_{l,ij}\geq 0 \\ 
\lambda_{l}(s_{j}), &  \mathrm{otherwise}
\end{array}} \right.
\end{equation}
where $s_i = \left \{ s_{l,i} \right \}_{l \in \left \{ 1,...,n_p \right \}}$ denotes the saturations of cell $i$.

The total face transmissibility $T_{ij}$ combines two half-transmissibilities in a half of the harmonic average
\begin{equation}
T_{ij} = \frac{T_i T_j}{T_i + T_j} \, , \qquad T_{i} = \frac{k_i A_{ij}}{d_i}.
\end{equation}
where $A_{ij}$ denotes the interface area, $k_i$ is the permeability of cell $i$, and $d_i$ is the length from the cell centroid to the interface.

In the finite volume formulation, the discrete source (or sink) term for a mesh cell containing a well (referred to as well cell) is written as (Peaceman 1983)
\begin{equation} 
Q_{l,i} = \textrm{WI}_i \, \left ( \rho_{l} \lambda_{l} \right )_i \left ( p_l - p^W \right )_i,
\end{equation}
which represents the well flux for phase $l$ in cell $i$. $p_{l,i}$ is well-cell pressure, $p^W_i$ is wellbore pressure, and $\textrm{WI}_i$ is well index.

The resulting discretized nonlinear system is written in residual format as
\begin{equation} 
\mathcal{R}(\textbf{\textit{u}}^{n+1}) = 0 
\end{equation}
where $\textbf{\textit{u}}$ represents the state variables (pressure and saturation) of mesh cells. The nonlinear system is often solved using the Newton method, which performs multiple iterations until convergence. For each timestep, with the solution $\textbf{\textit{u}}^{n}$, and a chosen timestep size $\Delta t$, the new state $\textbf{\textit{u}}^{n+1}$ is obtained.

\section{Surrogate models}

A simulator $\mathbb{H}$ maps the current state of mesh cells to the next timestep state. We denote a simulated rollout trajectory as $\left ( \textbf{\textit{u}}^{0}, \textbf{\textit{u}}^{1}, ..., \textbf{\textit{u}}^{n_t} \right )$, which is computed iteratively by $\textbf{\textit{u}}^{n+1} = \mathbb{H} \left ( \textbf{\textit{u}}^{n} \right )$ over $n_t$ timesteps.

The goal of our surrogate learning task is to replace the computationally expensive high-fidelity simulator with surrogate simulators that predict the next state 
\begin{equation} 
\textbf{\textit{u}}^{n+1} \approx \widehat{\textbf{\textit{u}}}^{n+1} = \mathbb{N} \left ( \textbf{\textit{u}}^{n}; \Theta \right ),
\end{equation}
where $\mathbb{N}$ is a next-step prediction model based on GNN, whose parameters $\Theta$ can be optimized for some end-to-end training objectives. $\widehat{\textbf{\textit{u}}}^{n+1}$ indicates the predicted state from the surrogate model. Given the initial state $\textbf{\textit{u}}^{0}$, $\mathbb{N} \left ( ; \Theta \right )$ can rapidly produce a rollout trajectory of states $\left ( \textbf{\textit{u}}^{0}, \widehat{\textbf{\textit{u}}}^{1}, ..., \widehat{\textbf{\textit{u}}}^{n_t} \right )$ in an autoregressive way.

The coupled multi-phase flow system (Equation (\ref{eq:mass_con})) has an intricate mixture of elliptic and hyperbolic characteristics. It is beneficial to employ specialized GNN architectures suitable for the specific characteristics of the coupled system. Therefore in the present work we seperately design and train two models that compute the solutions of pressure and saturation in a sequential manner as
\begin{equation}
\label{eq:n_PS}
\left\{
\begin{aligned}
\widehat{\textbf{\textit{p}}}^{n+1} & = \mathbb{N}_p \left ( \textbf{\textit{p}}^{n}, \textbf{\textit{s}}^{n}; \Theta_p \right ) , \\ 
\widehat{\textbf{\textit{s}}}^{n+1} & = \mathbb{N}_s \left ( \widehat{\textbf{\textit{p}}}^{n+1}, \textbf{\textit{s}}^{n}; \Theta_s \right ) .
\end{aligned}
\right.
\end{equation}
where $\mathbb{N}_p$ and $\mathbb{N}_s$ represent respectively the pressure and saturation models. At each time step, the saturation model takes the input from the pressure model.

The above process can be written using a compact operator as
\begin{equation} 
\left [ \widehat{\textbf{\textit{p}}}^{n+1}, \widehat{\textbf{\textit{s}}}^{n+1} \right ] = \mathbb{N}_s  \circ \mathbb{N}_p \left ( \textbf{\textit{p}}^{n}, \textbf{\textit{s}}^{n}; \Theta_p, \Theta_s \right ).
\label{eq:surrogate_model}
\end{equation}

\section{Graph Neural Networks}

We leverage the power of GNNs to construct data-driven surrogate simulators to approximate the PDE solutions (Equation (\ref{eq:surrogate_model})). GNNs provide a flexible and efficient way to operate over data that are structured as graphs, naturally fitting mesh-based simulations (Pfaff et al. 2020; Pilva and Zareei 2022).

\begin{figure}[!htb]
\centering
\includegraphics[scale=1.2]{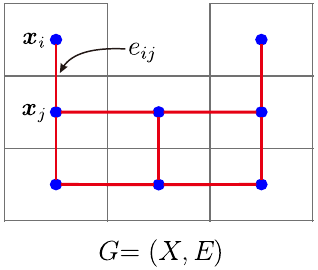}
\caption{Schematic of the graph representation of a simulation mesh. Blue dots indicate nodes (cell centroids), and red segments indicate edges (cell connections).}
\label{fig:graph_representation}
\end{figure}

A simulation mesh can be represented as the graph $G = (X, E)$ (\textbf{Fig.~\ref{fig:graph_representation}}) with nodes $X$ (blue dots), undirected edges $E$ (red line segments), and an adjacency matrix comprising edge connectivity. Let $\textbf{\textit{x}}_i$ be the cell centroid, and $\varepsilon_{ij}$ represents the connecting neighboring cells at $\textbf{\textit{x}}_i$ and $\textbf{\textit{x}}_j$. $\mathcal{N}(i)$ is the set of adjacent nodes around node $i$. We further denote the node and edge features by $\textbf{\textit{h}}_i$ and $\textbf{\textit{e}}_{ij}$ respectively.

A GNN-based model consists of multiple neural network layers, each aiming to aggregate local information across the one-hop neighborhood of each node and pass on to the neighbors. Stacking $n_L$ GNN layers allows the network to build node representations from the $n_L$-hop neighborhood. The fundamental operation in GNNs is the message-passing procedure, which is generally formulated as (Gilmer et al. 2017)
\begin{equation}
{\textbf{\textit{h}}}'_i = \gamma \left ( \textbf{\textit{h}}_i, \underset{j\in\mathcal{N}(i)}{\bigoplus} \psi \left ( \textbf{\textit{h}}_i, \textbf{\textit{h}}_j, \textbf{\textit{e}}_{ij} \right ) \right ),
\end{equation}
where $\bigoplus$ denotes a differentiable, permutation-invariant function (e.g., summation, mean, or maximum), and $\gamma$ and $\psi$ are differentiable neural networks such as MultiLayer Perceptrons (MLPs). The updated features ${\textbf{\textit{h}}}'_i$ are obtained by applying nonlinear transformations to the central feature vector $\textbf{\textit{h}}_i$ and the feature vectors $\textbf{\textit{h}}_j$ for all nodes $j$ in the neighborhood of node $i$. Each subsequent message-passing layer contains a separate set of network parameters, and operates on the output of the previous layer.

The design of $\bigoplus$ and $\gamma$ is what mostly distinguishes one type of GNN from another. We will present several popular message-passing-based GCN operators in the following sections and study them in the context of multi-phase flow and transport in porous media.

\subsection{GraphConv}

In our work, we consider weighted graphs and employ the GCN operator, GraphConv, from Morris et al. (2019). The new node features of a network layer are updated as
\begin{equation}
{\textbf{\textit{h}}}'_i = \sigma \left ( \textbf{W}_1 \textbf{\textit{h}}_i + \textbf{W}_2 \! \sum_{j\in\mathcal{N}(i)} \! w_{ij} \, \textbf{\textit{h}}_j \right ),
\end{equation}
where $\textbf{W}_1$ and $\textbf{W}_2$ are learnable parameter matrices, $w_{ij}$ denotes the edge weight (one-dimensional edge feature), and $\sigma$ denotes a nonlinear activation function, e.g., ReLU or Tanh.

\subsection{GATConv}

The graph attentional (GAT) operator (Veličković et al. 2017) adopts a self-attention process (Bahdanau et al. 2014) into graph learning with considering edge features. The attention mechanism allows a network to assign varying weights to neighboring nodes, enabling it to focus on more important information within the data. In GAT, a shared linear transformation is first applied to every node. Then we perform self-attention using a shared attentional mechanism which assigns an unnormalized coefficient for every node pair $(j, i)$ as
\begin{equation}
\xi_{ij} = \text{LeakyReLU}\left(\mathbf{a}^{\top}[\mathbf{W} h_i \| \mathbf{W} h_j]\right)
\end{equation}
which specifies the importance of node $j$'s features to node $i$. $\mathbf{W}$ is a weight matrix, and $\Vert$ denotes vector concatenation. Here the attention mechanism is a single-layer feed-forward neural network, parametrised by a weight vector $\mathbf{a}$, and followed by the LeakyReLU nonlinearity with negative slope 0.2.

To incorporate multi-dimensional edge features $\textbf{\textit{e}}_{ij}$, $\xi_{ij}$ may be computed as
\begin{equation}
\xi_{ij} = \text{LeakyReLU}\left(\mathbf{a}^{\top}[\mathbf{W}_1 h_i \| \mathbf{W}_1 h_j \| \mathbf{W}_2 \textbf{\textit{e}}_{ij}]\right)
\end{equation}
These coefficients are then normalised through softmax, in order to be comparable over different neighborhoods
\begin{equation}
\alpha_{ij} = \textrm{softmax}_j(\xi_{ij}) = \frac{\textrm{exp} \left ( \xi_{ij} \right )}{\sum_{k \in \mathcal{N}(i)\cup\left \{ i \right \}} \textrm{exp} \left ( \xi_{ik} \right ) }
\end{equation}

The attention scores are finally expressed as
\begin{equation}
\alpha_{ij} = \frac{\textrm{exp} \left ( \text{LeakyReLU}\left(\mathbf{a}^{\top}[\mathbf{W}_1 h_i \| \mathbf{W}_1 h_j \| \mathbf{W}_2 \textbf{\textit{e}}_{ij}]\right) \right )}{\sum_{k \in \mathcal{N}(i)\cup\left \{ i \right \}} \textrm{exp} \left ( \text{LeakyReLU}\left(\mathbf{a}^{\top}[\mathbf{W}_1 h_i \| \mathbf{W}_1 h_k \| \mathbf{W}_2 \textbf{\textit{e}}_{ik}]\right) \right ) }
\end{equation}

After obtaining the attention scores, we can update the layer embedding as a weighted sum of the transformed features
\begin{equation}
{\textbf{\textit{h}}}'_i = \sigma \left ( \alpha_{ii} \textbf{W}_1 \textbf{\textit{h}}_i + \! \sum_{j\in\mathcal{N}(i)} \! \alpha_{ij} \textbf{W}_1 \textbf{\textit{h}}_j \right )
\end{equation}

\begin{figure}[!htb]
\centering
\includegraphics[scale=0.52]{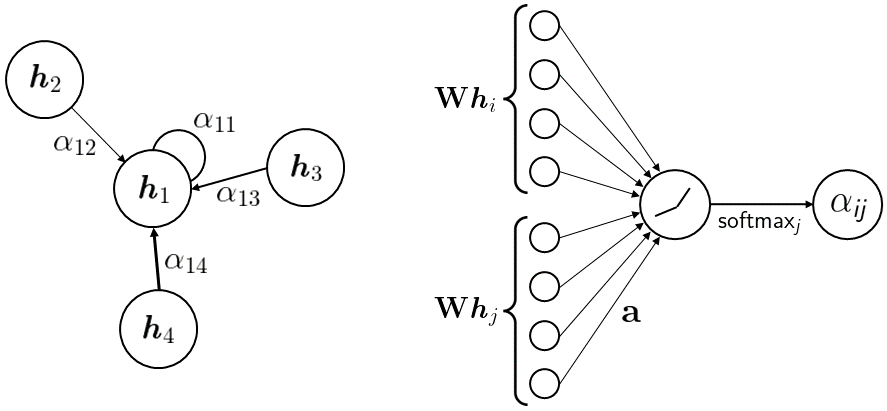}
\caption{Illustration of the GAT operator. Left: the importance of the neighboring nodes to node 1; Right: the attention scores from the attention mechanism.}
\label{fig:GAT_plot}
\end{figure}

An illustration of the GAT operator is shown in \textbf{Fig.~\ref{fig:GAT_plot}}.

\subsection{EdgeConv}

We additionally employ the edge convolution operator, EdgeConv from Wang et al. (2019). EdgeConv exploits local geometric structures by performing convolution operations on the edges linking neighboring node pairs. Here we choose an asymmetric edge function and then the layer output can be computed by
\begin{equation}
{\textbf{\textit{h}}}'_i = \underset{j\in\mathcal{N}(i)}{\textrm{max}} \Psi \! \left ( \textbf{\textit{h}}_i, \textbf{\textit{h}}_j - \textbf{\textit{h}}_i \right ).
\label{eq:edgeconv}
\end{equation}
where $\Psi$ denotes an MLP. As can be seen, the max aggregation operation is used on the generated edge features of all the edges emanating from a node. We find that EdgeConv can greatly improve the accuracy of the computed transport (saturation) dynamics. One possible reason is that the solution differences $(\textbf{\textit{h}}_j - \textbf{\textit{h}}_i)$ in the message passing mimic the differential operators of the discrete PDE system.

\section{Model architectures}

\begin{figure}[!htb]
\centering
\includegraphics[scale=0.71]{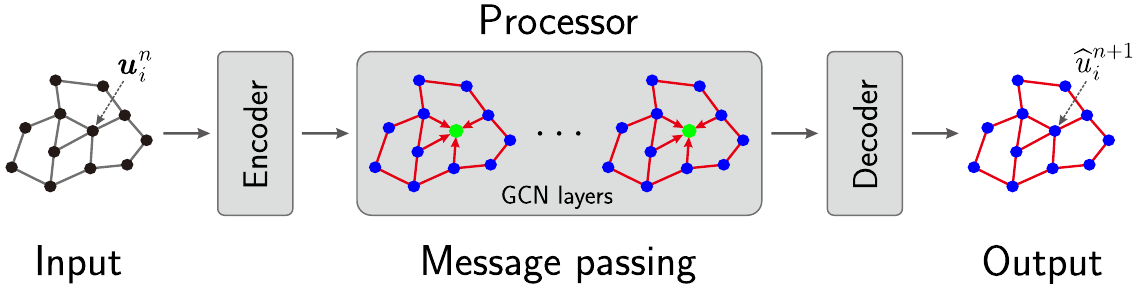}
\caption{Schematic of a general GCN model architecture.}
\label{fig:model_architecture}
\end{figure}

In this section, we present the detailed surrogate models which can predict the next-step dynamic states of the coupled PDE system. Our GCN models have an Encoder-Processor-Decoder structure. Schematic of a general GCN model architecture is plotted in \textbf{Fig.~\ref{fig:model_architecture}}. The initial node features are first encoded into latent vectors of size $n_{H}$. The input features $\textbf{\textit{h}}_i$ of mesh node $i$ for each timestep contain the dynamic variables (pressure and saturation), permeability, and pore volume. A one-hot vector indicating node type (distinguishing reservoir, production, and injection nodes), along with the well index are also added. In the GATConv operator, the transmissibility $T_{ij}$ of each connection and the vector of relative node positions $(\textbf{\textit{x}}_{ij} = \textbf{\textit{x}}_i - \textbf{\textit{x}}_j)$ are encoded as the multi-dimensional edge features. The relative positional information enables GNNs to possess spatial equivariance, which can be important for correctly capturing the directional flow patterns in the transport solutions (Iakovlev et al. 2020). For GraphConv, only $T_{ij}$ is taken as the edge feature (edge weight). Each feature is scaled individually to $[0, 1]$ using the min-max normalization method. The Decoder extracts one target output state (either $\widehat{\textbf{\textit{p}}}^{n+1}$ or $\widehat{\textbf{\textit{s}}}^{n+1}$) from the latent node features after the final processing layer. The Encoder and Decoder are two-layer MLPs with ReLU nonlinearities except for the output layer of the Decoder, after which we do not apply any nonlinearity.

The Processor of the pressure model $\mathbb{N}_p$ is constructed by stacking 8 identical GATConv layers with the mean aggregation operation and ReLU nonlinearities, to obtain a sequence of updated latent features. For the $\mathbb{N}_s$ model, we propose a combined architecture (3 EdgeConv followed by 5 GraphConv or GATConv layers with max aggregation), which is found to be quite effective for capturing the hyperbolic (saturation) solution. The Tanh activation function is applied. The sizes of hidden units for $\mathbb{N}_p$ and $\mathbb{N}_s$ are $n_{Hp} = 32$ and $n_{Hs} = 128$, respectively.

\section{Training procedure}

We train the GCN models using the dynamic state pairs $\left ( \textbf{\textit{u}}^{n}; \textbf{\textit{u}}^{n+1} \right )$ from $n_Y$ number of simulated rollout trajectories. We employ a mean squared error (MSE) loss between predictions $\widehat{\textbf{\textit{u}}}_y^{n+1}$ and their corresponding ground truth values $\textbf{\textit{u}}_y^{n+1}$ (simulator reference). The $L_2$ loss function is minimized through
\begin{equation}
\Theta^{*} = \underset{\Theta}{\textrm{argmin}} \, \frac{1}{n_Y} \frac{1}{n_t} \sum_{y=1}^{n_Y} \sum_{n=0}^{n_t-1} \left \| \widehat{\textbf{\textit{u}}}_y^{n+1} - \textbf{\textit{u}}_y^{n+1} \right \|_2^2
\end{equation}
where $n_t$ is the number of timesteps (temporal snapshots), and $\textbf{\textit{u}}_y^{n+1}$ denotes either pressure or water saturation of every mesh node, at time $t_{n+1}$, for training sample $y$. During training, the network weights of GNN are updated based on the gradient to the loss function through back-propagation.

Modeling a complex time-dependent PDE system requires the model to mitigate error accumulation over long rollout trajectories (Sanchez-Gonzalez et al. 2020). Because we only train our surrogates on ground-truth one-step data, we corrupt the input saturation states $\textbf{\textit{u}}_y^{n}$ with normal noise $N_s \left ( 0, \sigma_s = 0.02 \right )$ of zero mean and fixed variance. In this way, the rollouts of multiple timesteps from trained models become robust to their own noisy, previous predictions as input.

\section{Surrogate model evaluations}

We explore the prediction performance of the surrogate models and their generalization capabilities on out-of-training domain shapes and meshes. As an example, we consider 2D reservoir models in the x-z domain containing two wells (one injector and one producer) that operate under constant bottom-hole pressure (BHP). No-flow boundary condition is specified at the reservoir boundaries. The set-up of the base model is summarized in Table \ref{tab:specification_m}. Total simulation time is 100 days, with a number of 20 timesteps. The migration of multi-phase fluid is governed by the complex interplay of viscous, capillary, and gravity forces. Quadratic relative permeabilities are used. The capillary pressure is computed through a simple linear function as
\begin{equation}
p_c(s^{\ast}) = p_e \, s^{\ast}
\end{equation}
where $p_e = 7.25 \, \textrm{psi}$ is the capillary entry pressure, and $s^{\ast}$ is the normalized (effective) water saturation
\begin{equation}
s^{\ast} = \frac{s_w - s_{wr}}{1 - s_{wr}}
\end{equation}

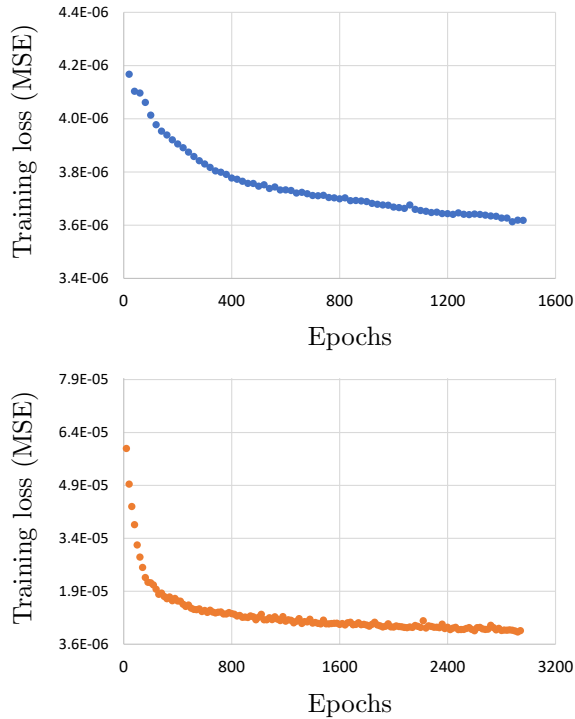
\begin{figure}[!htb]
\centering

\begin{minipage}{1\textwidth}
\centering
\begin{tikzpicture}
  \node (img)  {\includegraphics[scale=0.55]{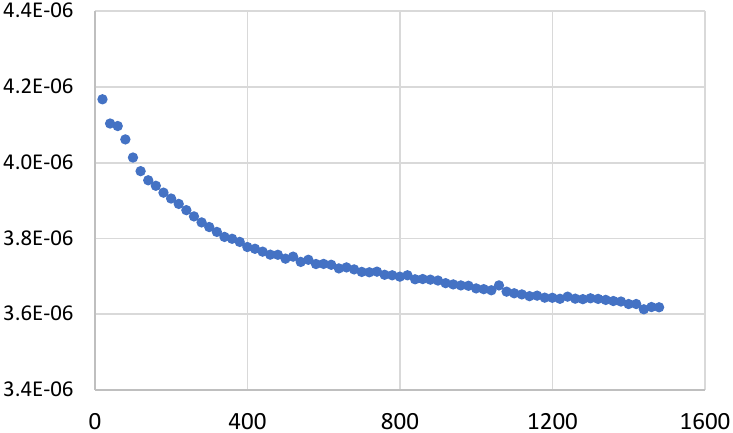}};
  \node[below=of img, node distance=0cm, xshift=0.45cm, yshift=1cm, font=\normalsize] {Epochs};
  \node[left=of img, node distance=0cm, rotate=90, anchor=center, xshift=0.2cm, yshift=-0.7cm, font=\normalsize] {Training loss (MSE)};
 \end{tikzpicture}
\end{minipage}%
\\
\begin{minipage}{1\textwidth}
\centering
\begin{tikzpicture}
  \node (img)  {\includegraphics[scale=0.55]{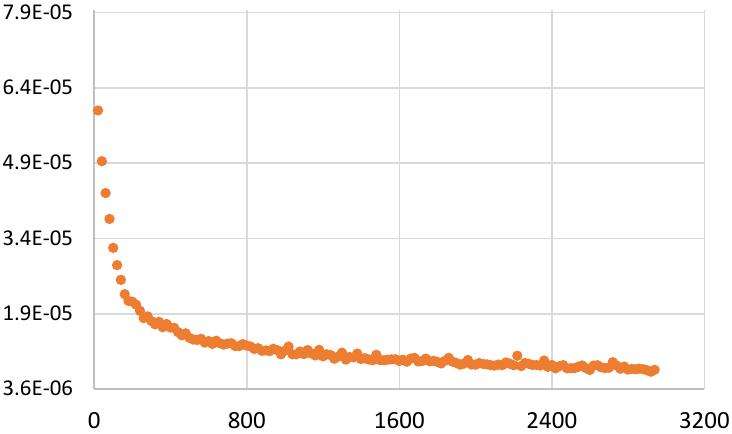}};
  \node[below=of img, node distance=0cm, xshift=0.45cm, yshift=1cm, font=\normalsize] {Epochs};
  \node[left=of img, node distance=0cm, rotate=90, anchor=center, xshift=0.2cm, yshift=-0.7cm, font=\normalsize] {Training loss (MSE)};
\end{tikzpicture}
\end{minipage}%

\caption{Training loss (MSE) curves of the pressure (upper) and saturation (lower) models.}
\label{fig:training_loss}
\end{figure}

There are a total of 180 high-fidelity simulation runs as training data samples with random well locations and rock properties on a regular $60 \times 60$ Cartesian mesh. The realizations of heterogeneous permeability and porosity fields are generated using a Gaussian distribution. The surrogate models have been trained on a NVIDIA Tesla V100 GPU using the Adam optimizer (Kingma and Ba 2014) with learning rate 1e-4. The training loss (MSE) curves are plotted in \textbf{Fig.~\ref{fig:training_loss}}.

It takes approximately 2 and 4 hours to train the pressure and saturation models, respectively. Note that the training times can be reduced by optimizing the hyperparameters of GCNs, and the learning rate schedule of the optimizer. Moreover, the large numbers of training epochs currently used are actually not necessary to reach reasonably low prediction errors. The trained models can predict a rollout trajectory in 0.05 seconds, achieving a significant reduction of computational time compared with the high-fidelity simulator. A single high-fidelity simulation run requires about 23 seconds on an Intel Core i7-12800HX CPU.

\begin{table}[!htb]
\centering
\caption{Set-up of the base model}
\label{tab:specification_m}
\begin{tabular}{|c|c|c|}
\hline
Parameter                  &  Value          & Unit   \\ \hline
Model sizes (x, y, z)      &  182, 3, 182    & m      \\ \hline
Initial pressure           &  2000           & psi    \\ \hline
Initial water saturation   &  0.01           & (-)       \\ \hline
Water density              &  1000            & $\textrm{kg}/\textrm{m}^3$  \\ \hline
Non-wetting phase reference density  &  800             & $\textrm{kg}/\textrm{m}^3$  \\ \hline
Water viscosity            &  1.0            & cP     \\ \hline
Non-wetting phase viscosity              &  2.0            & cP     \\ \hline
Rock compressibility       &  1e-8           & 1/bar  \\ \hline
Non-wetting phase compressibility        &  1e-4           & 1/bar  \\ \hline
Production BHP             &  1800           & psi  \\ \hline
Injection BHP              &  2200           & psi  \\ \hline
Total simulation time      &  100            & day     \\ \hline
Timestep size              &  5              & day     \\ \hline
\end{tabular}
\end{table}

\subsection{Regular Cartesian mesh}

We generate a total of 20 testing samples with random well locations and rock fields on the regular Cartesian mesh. Only qualitative and quantitative comparisons of the solutions (pressure and water saturation) at the end of the simulation between the surrogate (prediction) and high-fidelity (ground truth) simulators are analyzed, because the final snapshots of a rollout trajectory should exhibit the largest accumulated errors. The saturation results are compared using three models with different choices of convolution layers in the Processor of GCN: GATConv, EdgeConv+GraphConv, and EdgeConv+GATConv. The (GATConv) model (without EdgeConv) is treated as a baseline to demonstrate the relative performance of our new GCN architectures incorporating EdgeConv (\ref{eq:edgeconv}). The details of the saturation models are summarized in Table \ref{tab:model_details}.

\begin{table}[!htb]
\centering
\caption{Set-up of the saturation models.}
\label{tab:model_details}
\centering
\resizebox{\columnwidth}{!}
{
\begin{tabular}{|c|c|c|c|}
\hline
              & GATConv                 & EdgeConv+GraphConv                                                            & EdgeConv+GATConv                                                            \\ \hline
Processor     & Only GATConv layers     & \begin{tabular}[c]{@{}c@{}}3 EdgeConv \\ followed by 5 GraphConv\end{tabular} & \begin{tabular}[c]{@{}c@{}}3 EdgeConv \\ followed by 5 GATConv\end{tabular} \\ \hline
Edge features & $T_{ij}$ as edge weight & \begin{tabular}[c]{@{}c@{}}$T_{ij}$\\ $\textbf{\textit{x}}_{ij} = \textbf{\textit{x}}_i - \textbf{\textit{x}}_j$ \end{tabular}                          & \begin{tabular}[c]{@{}c@{}}$T_{ij}$\\ $\textbf{\textit{x}}_{ij} = \textbf{\textit{x}}_i - \textbf{\textit{x}}_j$ \end{tabular}                        \\ \hline
\end{tabular}
}
\end{table}

\begin{figure}[!htb]
\centering
\subfloat[Permeability (log)]{
\includegraphics[scale=0.38]{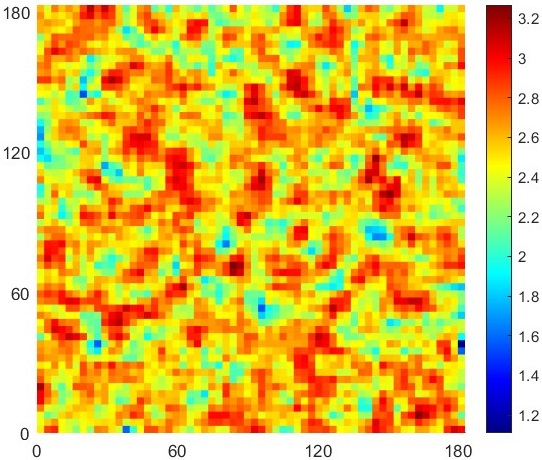}}
\
\subfloat[Pore volume]{
\includegraphics[scale=0.38]{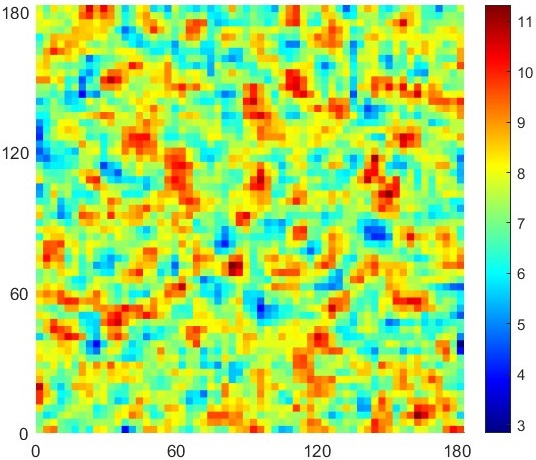}}
\caption{Permeability (md) and pore volume fields of Case 1 on the regular Cartesian mesh.}
\label{fig:regular_1_rock}
\end{figure}

\begin{figure}[!htb]
\centering
\subfloat[Pressure (psi)]{
\includegraphics[scale=0.49]{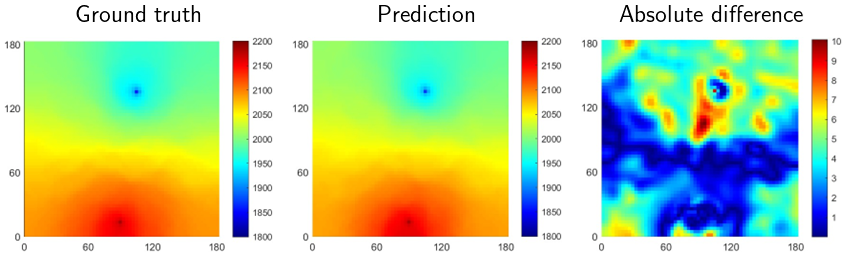}}
\
\subfloat[Saturation. First row: ground truth and prediction; Second row: absolute difference.]{
\includegraphics[scale=0.47]{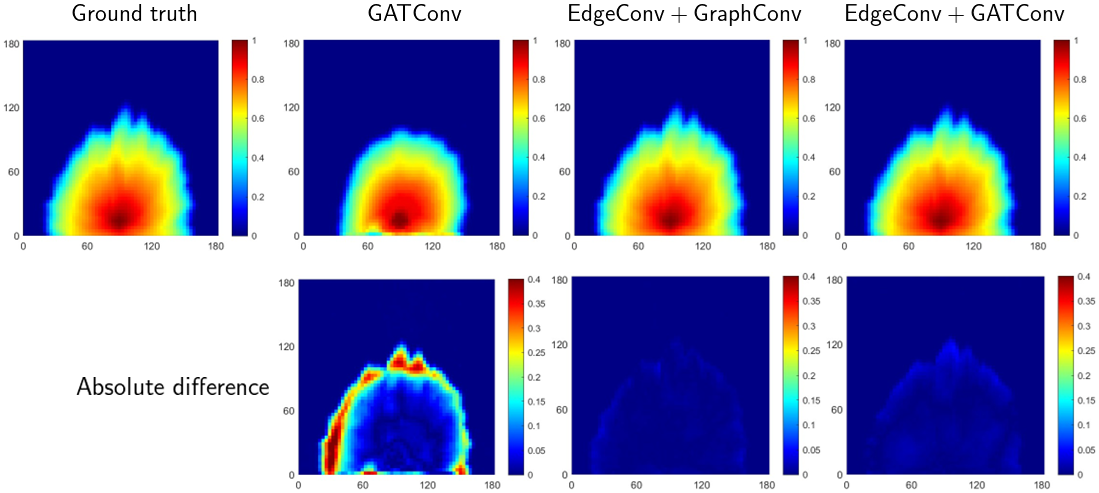}}
\caption{Solution profiles of Case 1 on the regular Cartesian mesh.}
\label{fig:regular_1_ps}
\end{figure}

\begin{figure}[!htb]
\centering
\subfloat[Permeability (log)]{
\includegraphics[scale=0.38]{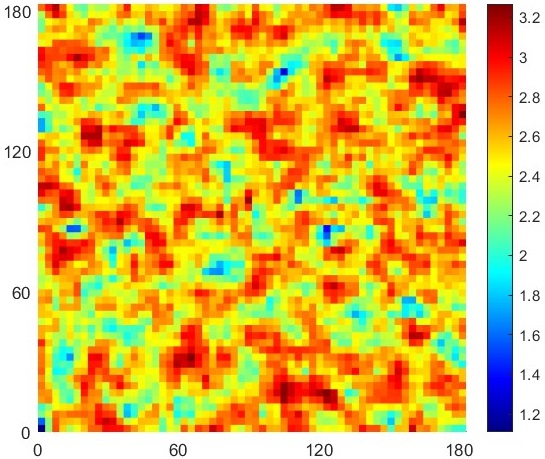}}
\
\subfloat[Pore volume]{
\includegraphics[scale=0.38]{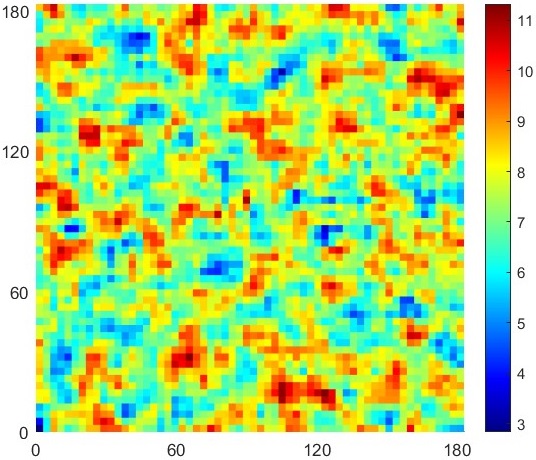}}
\caption{Permeability (md) and pore volume fields of Case 2 on the regular Cartesian mesh.}
\label{fig:regular_2_rock}
\end{figure}

\begin{figure}[!htb]
\centering
\subfloat[Pressure (psi)]{
\includegraphics[scale=0.49]{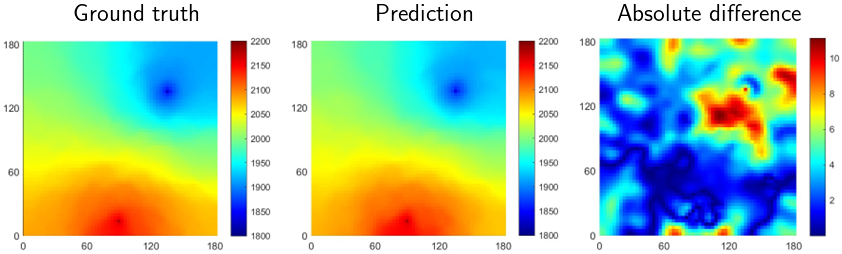}}
\
\subfloat[Saturation. First row: ground truth and prediction; Second row: absolute difference.]{
\includegraphics[scale=0.47]{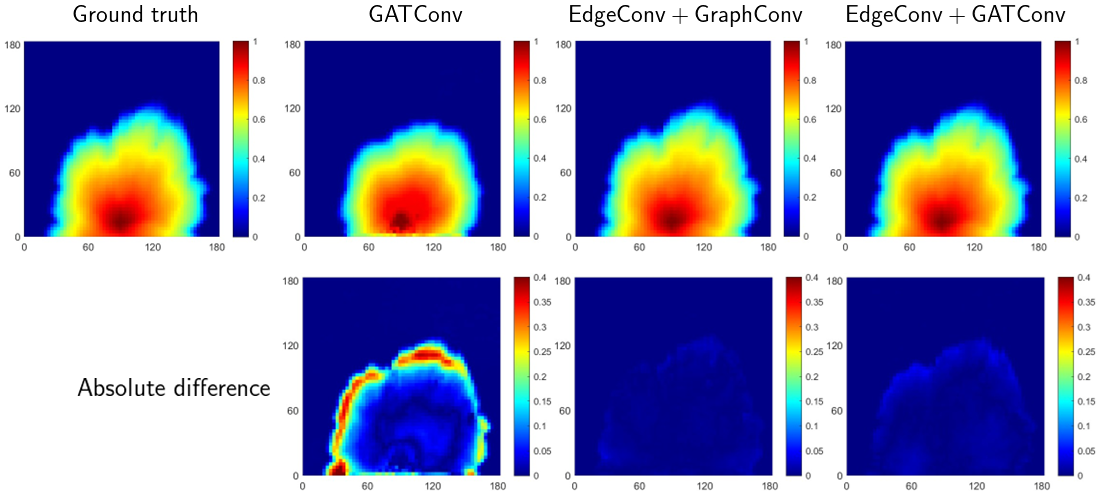}}
\caption{Solution profiles of Case 2 on the regular Cartesian mesh.}
\label{fig:regular_2_ps}
\end{figure}

\begin{figure}[!htb]
\centering
\subfloat[Permeability (log)]{
\includegraphics[scale=0.38]{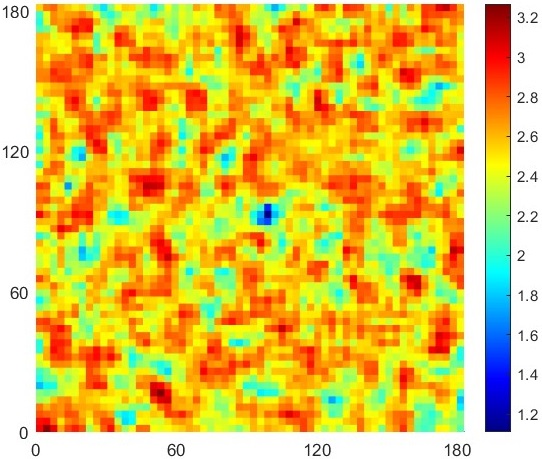}}
\
\subfloat[Pore volume]{
\includegraphics[scale=0.38]{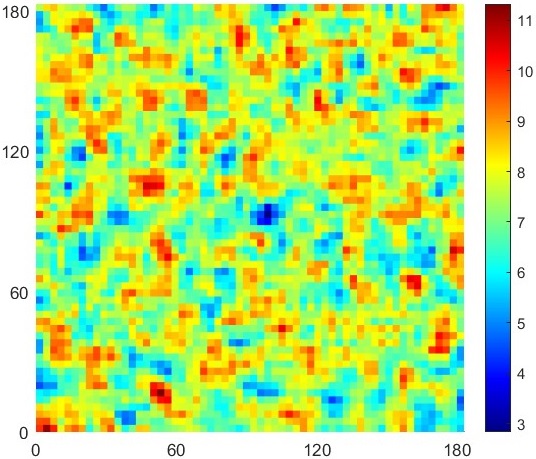}}
\caption{Permeability (md) and pore volume fields of Case 3 on the regular Cartesian mesh.}
\label{fig:regular_3_rock}
\end{figure}

\begin{figure}[!htb]
\centering
\subfloat[Pressure (psi)]{
\includegraphics[scale=0.49]{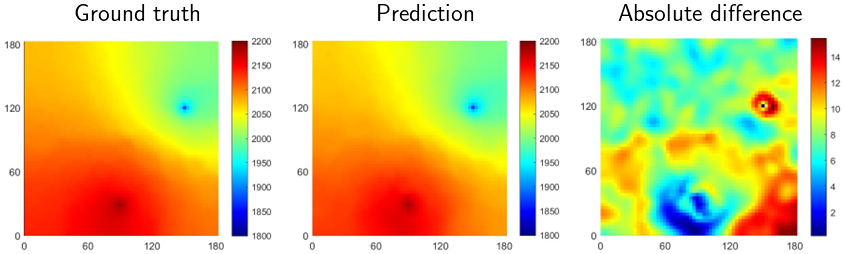}}
\
\subfloat[Saturation. First row: ground truth and prediction; Second row: absolute difference.]{
\includegraphics[scale=0.47]{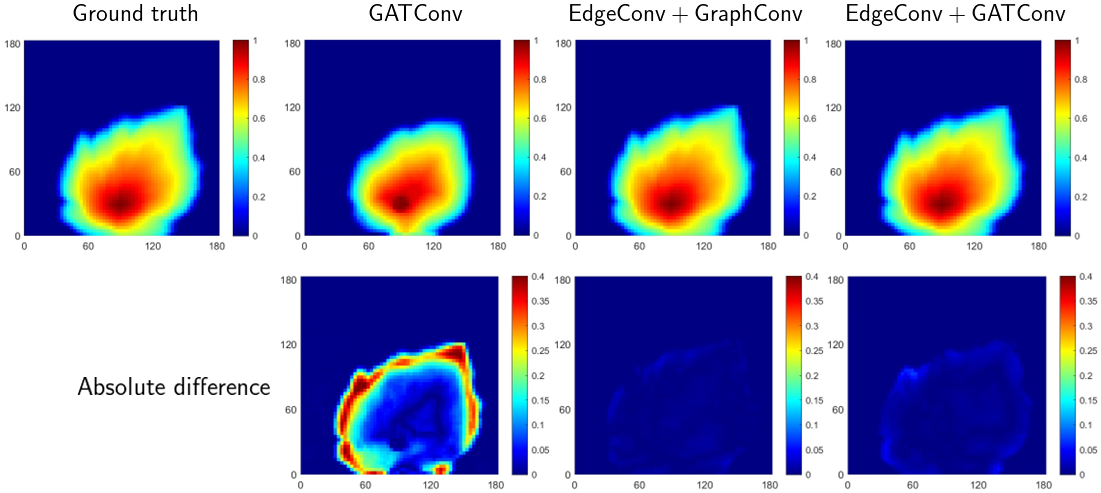}}
\caption{Solution profiles of Case 3 on the regular Cartesian mesh.}
\label{fig:regular_3_ps}
\end{figure}

We first present the predictions of three representative cases from the testing set. The rock fields of the three cases are shown in \textbf{Fig.~\ref{fig:regular_1_rock}}, \textbf{Fig.~\ref{fig:regular_2_rock}} and \textbf{Fig.~\ref{fig:regular_3_rock}}, respectively. The pressure and water saturation profiles are shown in \textbf{Fig.~\ref{fig:regular_1_ps}}, \textbf{Fig.~\ref{fig:regular_2_ps}} and \textbf{Fig.~\ref{fig:regular_3_ps}}, respectively. As can be seen, the pressure model is capable of providing physically smooth pressure solutions. 

The saturation fields are strongly impacted by the well locations and heterogeneous rock properties. We can see that the (GATConv) model cannot accurately predict the saturation solutions. Large saturation errors are evident near the water fronts. Moreover, some heterogeneous details inside the water plume are smeared out. In contrast, the two models with the combined architecture can reproduce both the shapes and heterogeneous details of the discontinuous saturation fronts quite well. This clearly demonstrates the benefit of EdgeConv for learning a hyperbolic system.

We further use two metrics to evaluate the accuracy of the predicted pressure and saturation snapshots (at the final time) for individual testing samples, including mean absolute error (MAE) and mean relative error (MRE) given as
\begin{equation}
\delta^{A} = \frac{1}{n_b} \sum_{i=1}^{n_b} \left \| \widehat{u}_i - u_i \right \|
\end{equation}

\begin{equation}
\delta_p^{R} = \frac{1}{n_b} \sum_{i=1}^{n_b} \frac{\left \| \widehat{p}_i - p_i \right \|}{p^0}
\end{equation}
where $n_b$ is the number of mesh cells in a snapshot, and $p^0$ is the initial reservoir pressure.

\begin{figure}[!htb]
\centering
\includegraphics[scale=0.46]{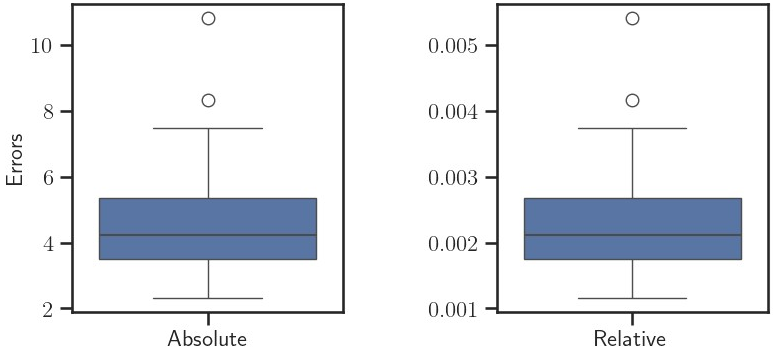}
\caption{Boxplots of the mean pressure errors over the testing cases on the regular Cartesian mesh.}
\label{fig:pres_error}
\end{figure}

\begin{figure}[!htb]
\centering
\includegraphics[scale=0.64]{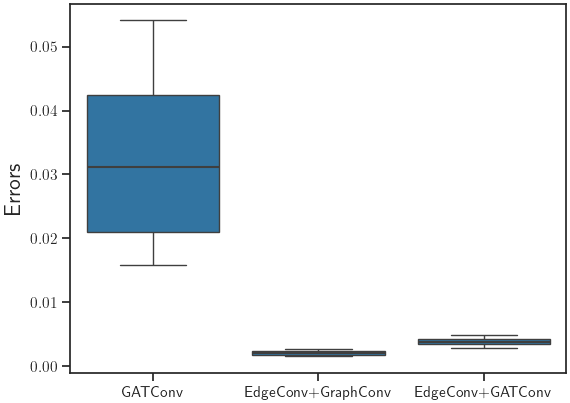}
\caption{Boxplots of the mean absolute saturation errors over the testing cases on the regular Cartesian mesh.}
\label{fig:sat_error}
\end{figure}

The boxplots of the error metrics $\delta_p$ and $\delta_s$ over all the testing samples are shown in \textbf{Fig.~\ref{fig:pres_error}} and \textbf{Fig.~\ref{fig:sat_error}}, respectively. We can see that the (EdgeConv+GraphConv) model performs slightly better than (EdgeConv+GATConv) in terms of the saturation predictions. Overall, the low mean errors indicate that the GCN-based surrogate models with EdgeConv accurately capture the evolutions of the pressure and saturation states in the full testing samples. This prediction capability is highly beneficial for a surrogate model in the context of uncertainty quantification. It is worth mentioning that even though our models were trained on the next-step predictions, the rollouts remain stable for multiple steps.

\subsection{Irregular Cartesian mesh}

\begin{figure}[!htb]
\centering
\subfloat[Permeability (log)]{
\includegraphics[scale=0.38]{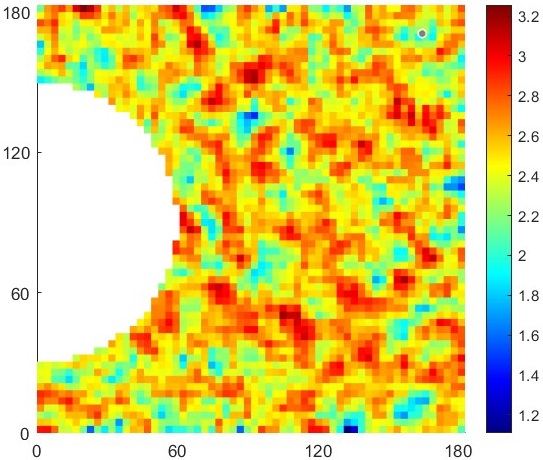}}
\
\subfloat[Pore volume]{
\includegraphics[scale=0.38]{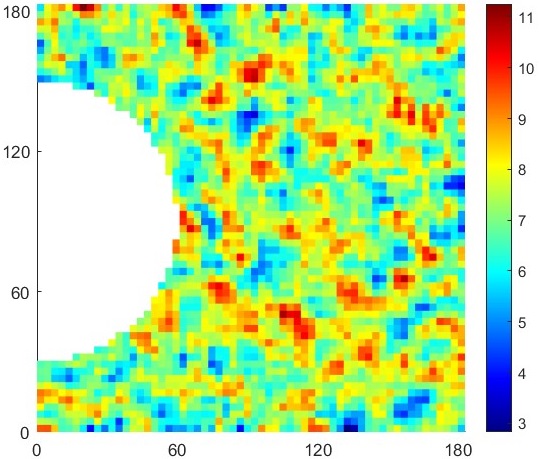}}
\caption{Permeability (md) and pore volume fields of Case 1 on irregular Cartesian mesh.}
\label{fig:irregular_1_rock}
\end{figure}

\begin{figure}[!htb]
\centering
\subfloat[Pressure (psi)]{
\includegraphics[scale=0.49]{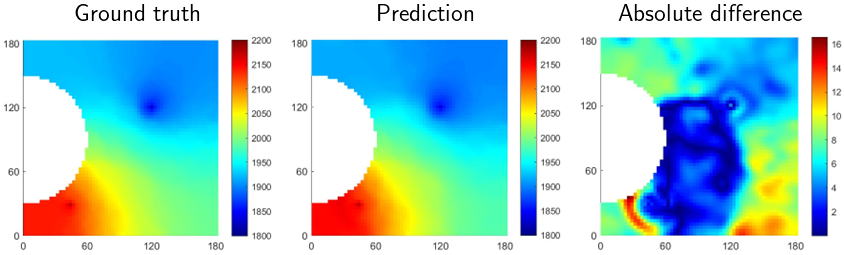}}
\
\subfloat[Saturation. First row: ground truth and prediction; Second row: absolute difference.]{
\includegraphics[scale=0.47]{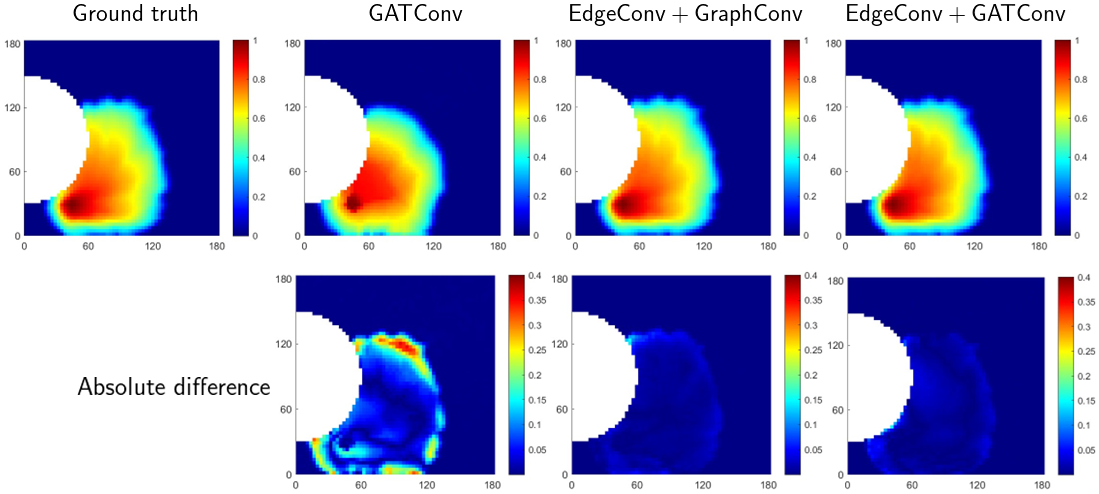}}
\caption{Solution profiles of Case 1 on irregular Cartesian mesh.}
\label{fig:irregular_1_ps}
\end{figure}

\begin{figure}[!htb]
\centering
\subfloat[Permeability (log)]{
\includegraphics[scale=0.38]{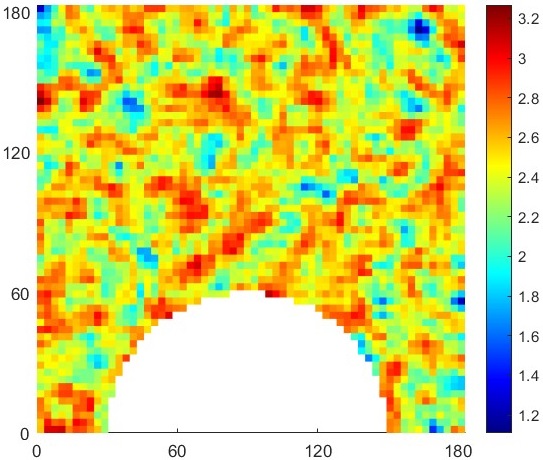}}
\
\subfloat[Pore volume]{
\includegraphics[scale=0.38]{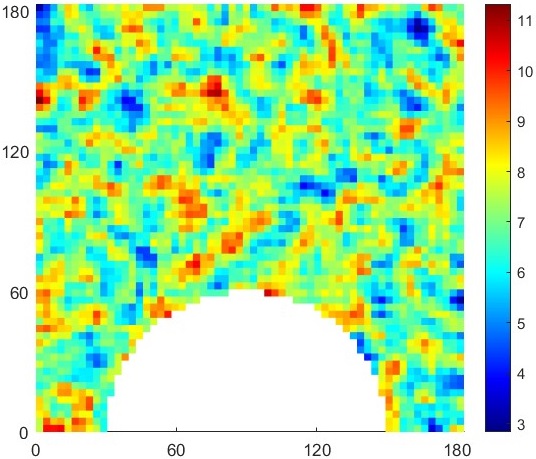}}
\caption{Permeability (md) and pore volume fields of Case 2 on irregular Cartesian mesh.}
\label{fig:irregular_2_rock}
\end{figure}

\begin{figure}[!htb]
\centering
\subfloat[Pressure (psi)]{
\includegraphics[scale=0.49]{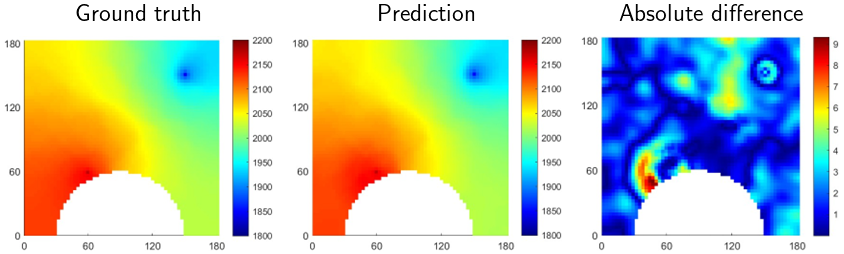}}
\
\subfloat[Saturation. First row: ground truth and prediction; Second row: absolute difference.]{
\includegraphics[scale=0.47]{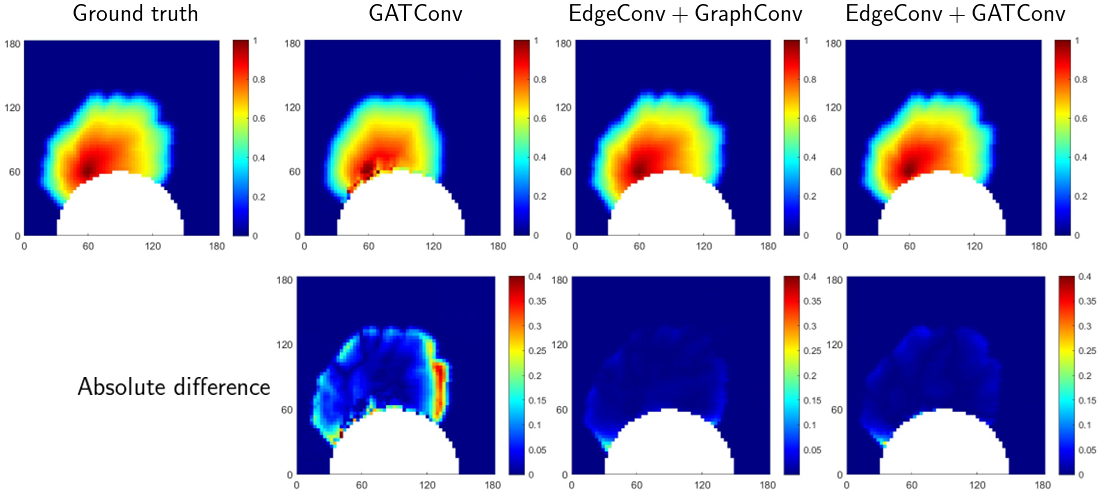}}
\caption{Solution profiles of Case 2 on irregular Cartesian mesh.}
\label{fig:irregular_2_ps}
\end{figure}

\begin{figure}[!htb]
\centering
\subfloat[Permeability (log)]{
\includegraphics[scale=0.38]{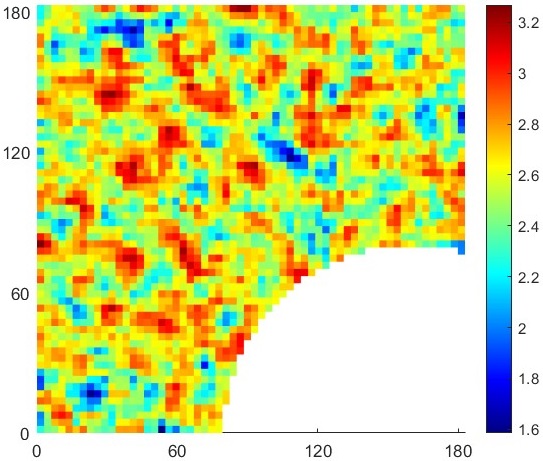}}
\
\subfloat[Pore volume]{
\includegraphics[scale=0.38]{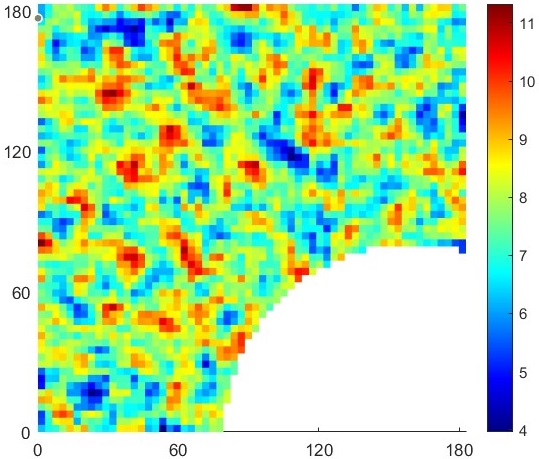}}
\caption{Permeability (md) and pore volume fields of Case 3 on irregular Cartesian mesh.}
\label{fig:irregular_3_rock}
\end{figure}

\begin{figure}[!htb]
\centering
\subfloat[Pressure (psi)]{
\includegraphics[scale=0.49]{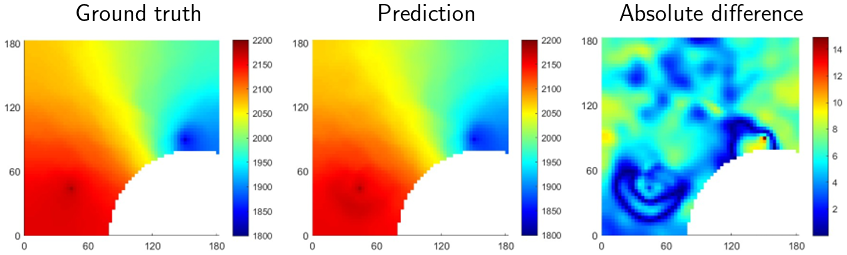}}
\
\subfloat[Saturation. First row: ground truth and prediction; Second row: absolute difference.]{
\includegraphics[scale=0.47]{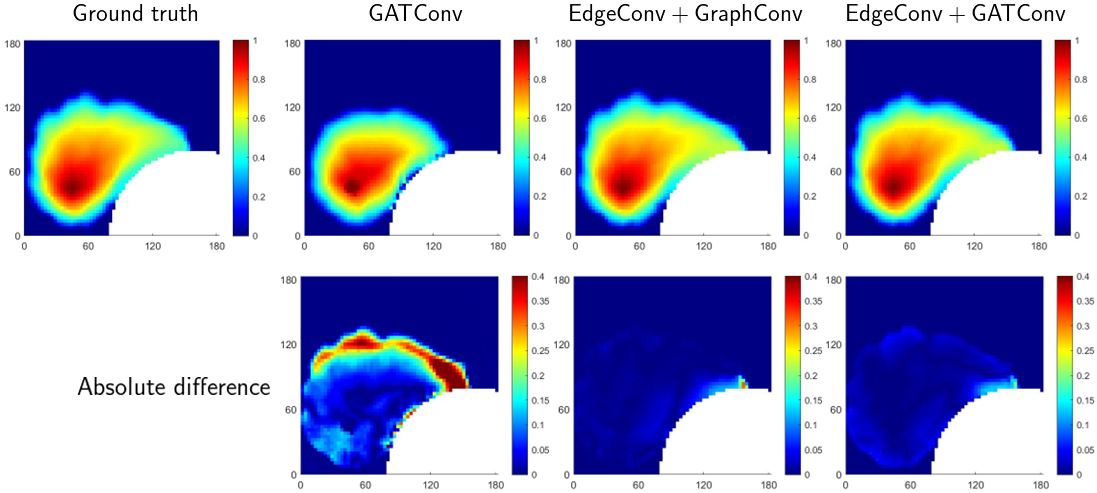}}
\caption{Solution profiles of Case 3 on irregular Cartesian mesh.}
\label{fig:irregular_3_ps}
\end{figure}

We further evaluate the generalization ability of the trained surrogate simulators, using three test cases with irregular mesh geometries. The rock property fields of the three cases are shown in \textbf{Fig.~\ref{fig:irregular_1_rock}}, \textbf{Fig.~\ref{fig:irregular_2_rock}} and \textbf{Fig.~\ref{fig:irregular_3_rock}}, respectively. Note that each case has a different number of graph nodes due to the geometry of different shapes. The solution profiles of the three cases for different GCN models are shown in \textbf{Fig.~\ref{fig:irregular_1_ps}}, \textbf{Fig.~\ref{fig:irregular_2_ps}} and \textbf{Fig.~\ref{fig:irregular_3_ps}}, respectively.

Again we can see that the (GATConv) model produces large saturation errors in all the testing cases. In comparison, the surrogates with the combined architecture predict the state evolutions with high accuracy. Visually there is no significant saturation difference with the reference simulations, except within certain regions near the domain boundaries. The results demonstrate that our Graph Convolutional Network can generalize well to unseen domain geometries, even though our models were trained using only the data samples on a regular square domain.

\subsection{PEBI mesh}

\begin{figure}[!htb]
\centering
\subfloat[Permeability (log)]{
\includegraphics[scale=0.32]{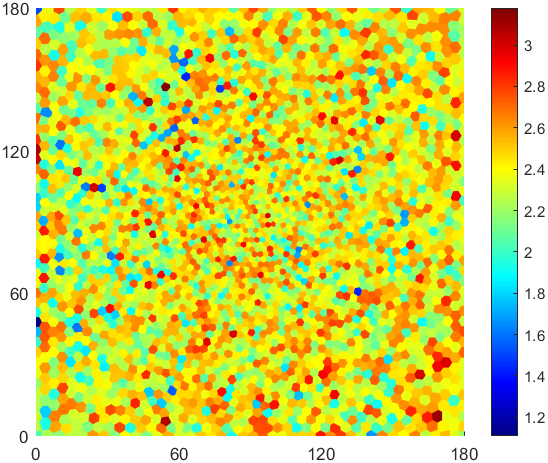}}
\
\subfloat[Pore volume]{
\includegraphics[scale=0.32]{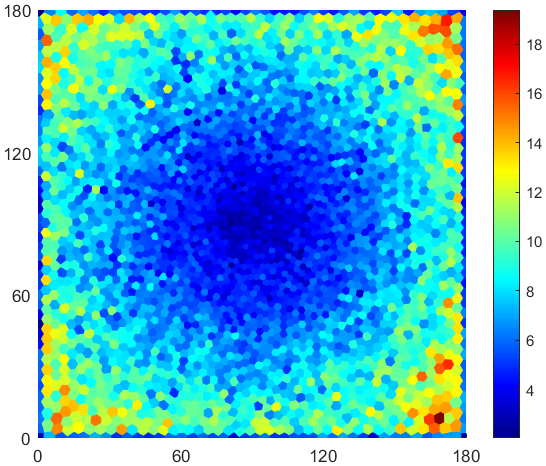}}
\caption{Permeability (md) and pore volume fields of Case 1 on the PEBI mesh.}
\label{fig:PEBI_1_rock}
\end{figure}

\begin{figure}[!htb]
\centering
\subfloat[Pressure (psi)]{
\includegraphics[scale=0.49]{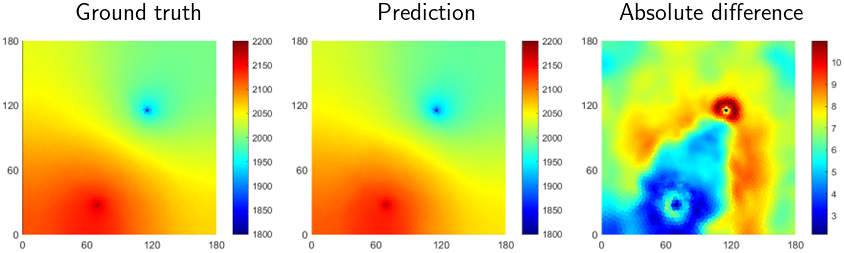}}
\
\subfloat[Saturation]{
\includegraphics[scale=0.49]{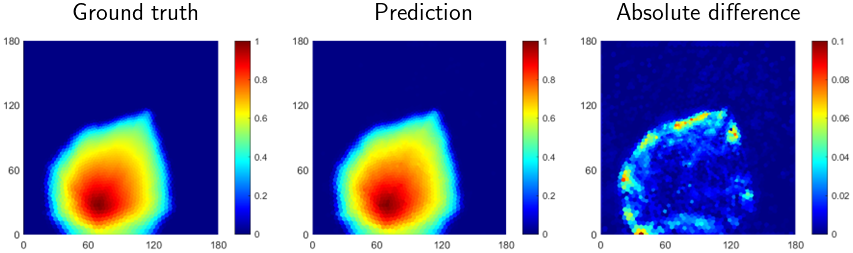}}
\caption{Solution profiles of Case 1 on the PEBI mesh.}
\label{fig:PEBI_1_ps}
\end{figure}

\begin{figure}[!htb]
\centering
\subfloat[Permeability (log)]{
\includegraphics[scale=0.32]{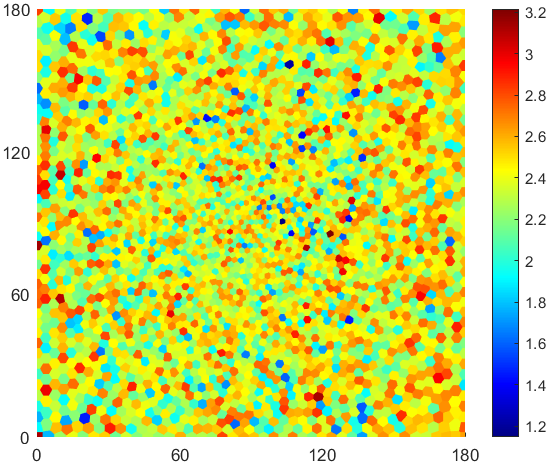}}
\
\subfloat[Pore volume]{
\includegraphics[scale=0.32]{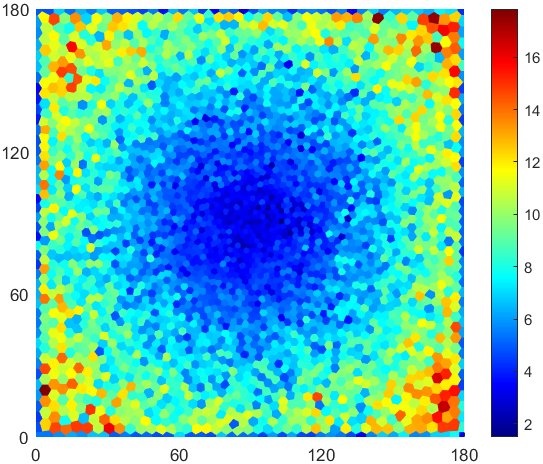}}
\caption{Permeability (md) and pore volume fields of Case 2 on the PEBI mesh.}
\label{fig:PEBI_2_rock}
\end{figure}

\begin{figure}[!htb]
\centering
\subfloat[Pressure (psi)]{
\includegraphics[scale=0.49]{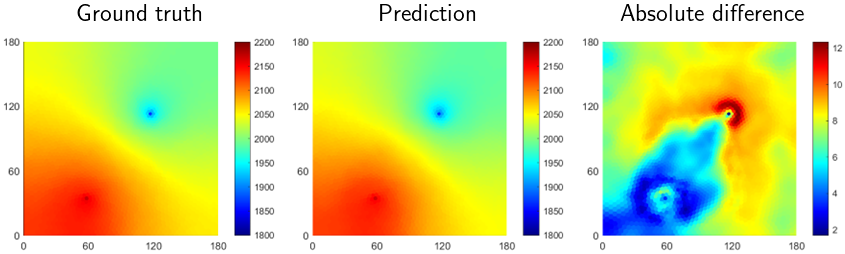}}
\
\subfloat[Saturation]{
\includegraphics[scale=0.49]{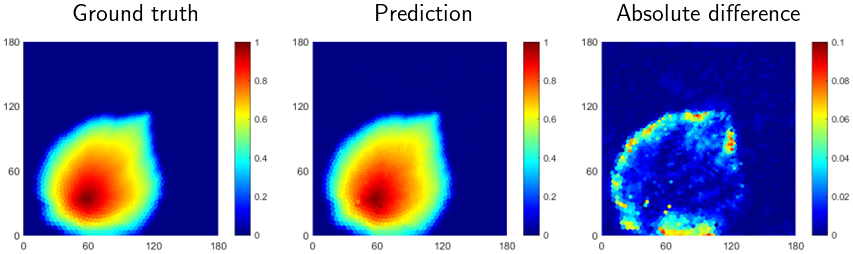}}
\caption{Solution profiles of Case 2 on the PEBI mesh.}
\label{fig:PEBI_2_ps}
\end{figure}

\begin{figure}[!htb]
\centering
\subfloat[Permeability (log)]{
\includegraphics[scale=0.32]{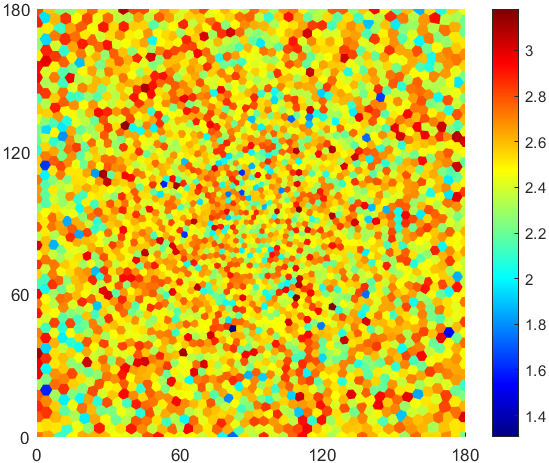}}
\
\subfloat[Pore volume]{
\includegraphics[scale=0.32]{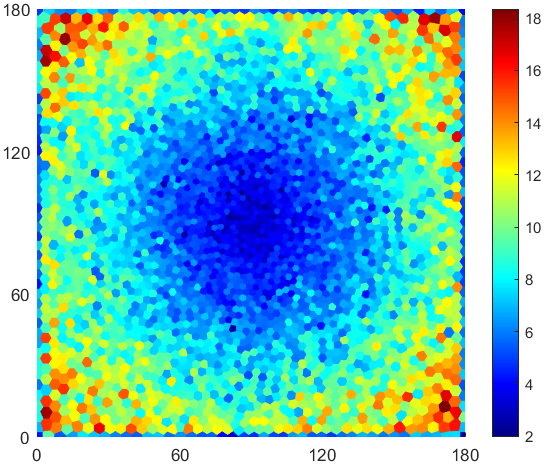}}
\caption{Permeability (md) and pore volume fields of Case 3 on the PEBI mesh.}
\label{fig:PEBI_3_rock}
\end{figure}

\begin{figure}[!htb]
\centering
\subfloat[Pressure (psi)]{
\includegraphics[scale=0.49]{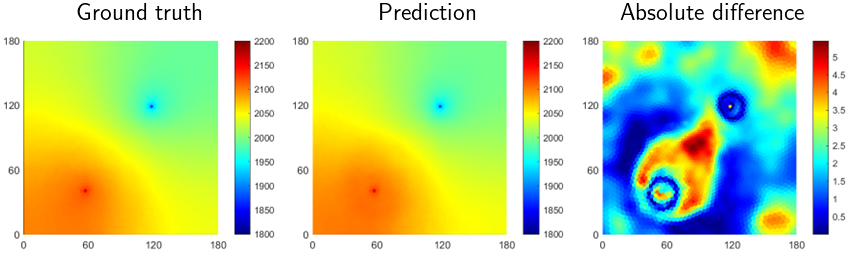}}
\
\subfloat[Saturation]{
\includegraphics[scale=0.49]{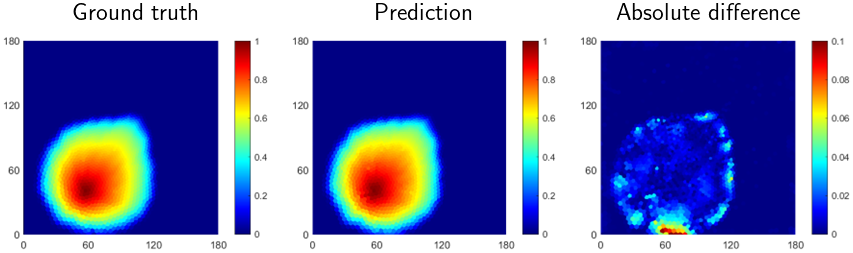}}
\caption{Solution profiles of Case 3 on the PEBI mesh.}
\label{fig:PEBI_3_ps}
\end{figure}

Furthermore, we perform testing on a perpendicular bisector (PEBI) mesh using the previously trained models. The (EdgeConv+GATConv) model is employed for saturation predictions. We present the predictions of three representative cases. The rock fields of the three cases are shown in \textbf{Fig.~\ref{fig:PEBI_1_rock}}, \textbf{Fig.~\ref{fig:PEBI_2_rock}} and \textbf{Fig.~\ref{fig:PEBI_3_rock}}, respectively. The pressure and water saturation profiles are shown in \textbf{Fig.~\ref{fig:PEBI_1_ps}}, \textbf{Fig.~\ref{fig:PEBI_2_ps}} and \textbf{Fig.~\ref{fig:PEBI_3_ps}}, respectively.

We can observe that the solutions from the surrogates closely match the high-fidelity simulations. Small saturation errors are, however, noticeable near the boundaries and fluid fronts. Our GCN models exhibit promising generalizability to unstructured meshes, suggesting that the networks learn a general understanding of the physical processes of the multi-phase flow and transport PDE system.

\section{Summary}

We apply GCNs for surrogate modeling to approximate the spatial-temporal solutions of multi-phase flow and transport in porous media. We propose a new GCN architecture suited to the hyperbolic character of the coupled PDE system, to better capture the saturation dynamics. Our surrogate models can offer significant speedups (460-fold) compared to the high-fidelity simulator for the conditions examined in the present study.

The prediction performance of the trained surrogates and their generalization capabilities on out-of-training domain shapes and meshes are evaluated using 2D heterogeneous test cases. The results show that our surrogates accurately predict the evolutions of the pressure and saturation states under the complex interplay of viscous, capillary, and gravitational forces. Even though the models were trained on the next-step predictions, the rollouts remain stable for multiple timesteps. The saturation model based on the GCN architecture incorporating EdgeConv can reproduce both the shapes and heterogeneous details of the discontinuous saturation fronts with high accuracy. Moreover, we demonstrate that the GCN-based models generalize well to unseen domain geometries and unstructured meshes.

Although the current examples are based on the simplified scenario of immiscible two-phase flow, our GCN framework can be readily applied to learn more complex processes such as 3D simulations of multicomponent (compositional) fluid systems.

\section*{Acknowledgements}

We thank Sidian Chen at The University of Arizona for constructive discussions.

\section*{References}

Brenier, Y. and Jaffré, J., 1991. Upstream differencing for multiphase flow in reservoir simulation. SIAM journal on numerical analysis, 28(3), pp.685-696.

Bahdanau, D., Cho, K. and Bengio, Y., 2014. Neural machine translation by jointly learning to align and translate. arXiv preprint arXiv:1409.0473.

Battaglia, P., Pascanu, R., Lai, M. and Jimenez Rezende, D., 2016. Interaction networks for learning about objects, relations and physics. Advances in neural information processing systems, 29.

Bar-Sinai, Y., Hoyer, S., Hickey, J. and Brenner, M.P., 2019. Learning data-driven discretizations for partial differential equations. Proceedings of the National Academy of Sciences, 116(31), pp.15344-15349.

Belbute-Peres, F.D.A., Economon, T. and Kolter, Z., 2020, November. Combining differentiable PDE solvers and graph neural networks for fluid flow prediction. In international conference on machine learning (pp. 2402-2411). PMLR.

Brandstetter, J., Worrall, D. and Welling, M., 2022. Message passing neural PDE solvers. arXiv preprint arXiv:2202.03376.

Chen, J., Hachem, E. and Viquerat, J., 2021. Graph neural networks for laminar flow prediction around random two-dimensional shapes. Physics of Fluids, 33(12), p.123607.

Guo, X., Li, W. and Iorio, F., 2016, August. Convolutional neural networks for steady flow approximation. In Proceedings of the 22nd ACM SIGKDD international conference on knowledge discovery and data mining (pp. 481-490).

Gilmer, J., Schoenholz, S.S., Riley, P.F., Vinyals, O. and Dahl, G.E., 2017, July. Neural message passing for quantum chemistry. In International conference on machine learning (pp. 1263-1272). PMLR.

Iakovlev, V., Heinonen, M. and Lähdesmäki, H., 2020. Learning continuous-time pdes from sparse data with graph neural networks. arXiv preprint arXiv:2006.08956.

Jiang, Z., Tahmasebi, P. and Mao, Z., 2021. Deep residual U-net convolution neural networks with autoregressive strategy for fluid flow predictions in large-scale geosystems. Advances in Water Resources, 150, p.103878.

Kingma, D.P. and Ba, J., 2014. Adam: A method for stochastic optimization. arXiv preprint arXiv:1412.6980.

Kipf, T.N. and Welling, M., 2016. Semi-supervised classification with graph convolutional networks. arXiv preprint arXiv:1609.02907.

Krizhevsky, A., Sutskever, I. and Hinton, G.E., 2017. Imagenet classification with deep convolutional neural networks. Communications of the ACM, 60(6), pp.84-90.

Kutz, J.N., 2017. Deep learning in fluid dynamics. Journal of Fluid Mechanics, 814, pp.1-4.

LeCun, Y., Bengio, Y. and Hinton, G., 2015. Deep learning. nature, 521(7553), pp.436-444.

Long, Z., Lu, Y., Ma, X. and Dong, B., 2018, July. Pde-net: Learning pdes from data. In International conference on machine learning (pp. 3208-3216). PMLR.

Morris, C., Ritzert, M., Fey, M., Hamilton, W.L., Lenssen, J.E., Rattan, G. and Grohe, M., 2019, July. Weisfeiler and leman go neural: Higher-order graph neural networks. In Proceedings of the AAAI conference on artificial intelligence (Vol. 33, No. 01, pp. 4602-4609).

Mo, S., Zhu, Y., Zabaras, N., Shi, X. and Wu, J., 2019. Deep convolutional encoder‐decoder networks for uncertainty quantification of dynamic multiphase flow in heterogeneous media. Water Resources Research, 55(1), pp.703-728.

Maldonado-Cruz, E. and Pyrcz, M.J., 2022. Fast evaluation of pressure and saturation predictions with a deep learning surrogate flow model. Journal of Petroleum Science and Engineering, 212, p.110244.

Maucec, M. and Jalali, R., 2022. GeoDIN-Geoscience-Based Deep Interaction Networks for Predicting Flow Dynamics in Reservoir Simulation Models. SPE Journal, 27(03), pp.1671-1689.

Peaceman, D.W., 1983. Interpretation of well-block pressures in numerical reservoir simulation with nonsquare grid blocks and anisotropic permeability. Society of Petroleum Engineers Journal, 23(03), pp.531-543.

Pfaff, T., Fortunato, M., Sanchez-Gonzalez, A. and Battaglia, P.W., 2020. Learning mesh-based simulation with graph networks. arXiv preprint arXiv:2010.03409.

Pilva, P. and Zareei, A., 2022. Learning time-dependent PDE solver using Message Passing Graph Neural Networks. arXiv preprint arXiv:2204.07651.

Sammon, P.H., 1988. An analysis of upstream differencing. SPE reservoir engineering, 3(03), pp.1-053.

Sanchez-Gonzalez, A., Godwin, J., Pfaff, T., Ying, R., Leskovec, J. and Battaglia, P., 2020, November. Learning to simulate complex physics with graph networks. In International conference on machine learning (pp. 8459-8468). PMLR.

Santos, J.E., Xu, D., Jo, H., Landry, C.J., Prodanović, M. and Pyrcz, M.J., 2020. PoreFlow-Net: A 3D convolutional neural network to predict fluid flow through porous media. Advances in Water Resources, 138, p.103539.

Tang, M., Liu, Y. and Durlofsky, L.J., 2020. A deep-learning-based surrogate model for data assimilation in dynamic subsurface flow problems. Journal of Computational Physics, 413, p.109456.

Veličković, P., Cucurull, G., Casanova, A., Romero, A., Lio, P. and Bengio, Y., 2017. Graph attention networks. arXiv preprint arXiv:1710.10903.

Vinuesa, R. and Brunton, S.L., 2022. Enhancing computational fluid dynamics with machine learning. Nature Computational Science, 2(6), pp.358-366.

Wang, Y., Sun, Y., Liu, Z., Sarma, S.E., Bronstein, M.M. and Solomon, J.M., 2019. Dynamic graph cnn for learning on point clouds. Acm Transactions On Graphics (tog), 38(5), pp.1-12.

Wang, Y. and Lin, G., 2020. Efficient deep learning techniques for multiphase flow simulation in heterogeneous porousc media. Journal of Computational Physics, 401, p.108968.

Wang, Y.D., Chung, T., Armstrong, R.T. and Mostaghimi, P., 2021. ML-LBM: Predicting and accelerating steady state flow simulation in porous media with convolutional neural networks. Transport in Porous Media, 138(1), pp.49-75.

Wen, G., Hay, C. and Benson, S.M., 2021. CCSNet: a deep learning modeling suite for CO2 storage. Advances in Water Resources, 155, p.104009.

Yan, B., Harp, D.R., Chen, B., Hoteit, H. and Pawar, R.J., 2022. A gradient-based deep neural network model for simulating multiphase flow in porous media. Journal of Computational Physics, 463, p.111277.

Zhang, K., Wang, Y., Li, G., Ma, X., Cui, S., Luo, Q., Wang, J., Yang, Y. and Yao, J., 2021. Prediction of field saturations using a fully convolutional network surrogate. SPE Journal, 26(04), pp.1824-1836.

\end{document}